\documentclass[12pt]{article}

\usepackage[margin=1in]{geometry} 
\usepackage{amsmath,amsthm,amssymb}
\usepackage{textcomp}
\usepackage[dvips,final, pdftex]{graphicx}         		
\usepackage[greek, english]{babel}                			
\usepackage[hang,bf]{caption}              			
\usepackage{rotating}                     		 		
\usepackage{subfigure}
\usepackage{units}
\usepackage{fancyhdr}
\usepackage{lscape}
\usepackage{amsfonts}
\usepackage{float}
\usepackage{graphicx}
\usepackage{longtable}
\usepackage{fancybox}
\usepackage[hidelinks]{hyperref}
\usepackage{here}
\usepackage{setspace} 
\usepackage{units} 
\usepackage{pdfpages}
\usepackage{lscape} 		
\usepackage{pdflscape}

\usepackage{upgreek}

\usepackage{listings}

\begin{document}
\title{Frictional half-plane contact problems subject to alternating normal and shear loads and tension in the steady state} 
\author{H. Andresen$^{\,\text{a,}}$\footnote{Corresponding author: \textit{Tel}.: +44 1865 273811; \newline \indent \indent \textit{E-mail address}: hendrik.andresen@eng.ox.ac.uk (H. Andresen).}$\,\,$, D.A. Hills$^{\,\text{a}}$, J.R. Barber$^{\,\text{b}}$, J.V\'azquez$^{\,\text{c}}$\\ \\
\scriptsize{$^{\text{a}}$ Department of Engineering Science, University of Oxford, Parks Road, OX1 3PJ Oxford, United Kingdom} \\
%\small{Email: hendrik.andresen@eng.ox.ac.uk, david.hills@eng.ox.ac.uk} \\
$\,$\scriptsize{$^{\text{b}}$ Department of Mechanical Engineering, University of Michigan, Ann Arbor, MI 48109-2125, USA}\\
$\,$\scriptsize{$^{\text{c}}$ Departamento de Ingenier\'ia Me\'canica y Fabricaci\'on, Universidad de Sevilla, Camino de los Descubrimientos,} \\ 
\scriptsize{41092 Sevilla, Spain}}
%\small{Email: jesusvaleo@us.es}
\date{}
\maketitle
\begin{center}
	\line(1,0){470}
\end{center}
\begin{abstract}
\footnotesize{	
The problem of a general, symmetric contact, between elastically similar
bodies, and capable of idealisation using half-plane theory, is studied in the
presence of interfacial friction. It is subject to a constant set of loads -
normal force, shear force and bulk tension parallel with the interface -
together with an oscillatory set of the same quantities, and is in the steady
state. Partial slip conditions are expected to ensue for a range of these
quantities, and the permanent stick zone is explicitly established, thereby
effectively specifying the maximum extent of the slip zones. Exact and
approximate, easy to apply recipes are obtained.\\}

  \noindent \scriptsize{\textit{Keywords}: Contact mechanics; Half-plane theory; Partial slip; Varying normal and shear loads; Bulk tension; Asymptotic methods}
  \end{abstract}
\begin{center}
	\line(1,0){470}
\end{center}

\section{Introduction}

\hspace{0.4cm} Fretting damage is a serious cause of crack nucleation and our
understanding of contacts suffering loading which
gives rise to partial slip remains incomplete. Our laboratory is currently
investigating a number of practical fretting problems in the aerospace,
automotive and subsea industries. There is a striking similarity between
some aspects of the first and last applications; a dovetail root of a gas
turbine fan-blade has, as each flank, a contact resembling closely a flat
punch with rounded edges, and the locking segments employed in wellhead
connectors are geometrically very similar. There is a further aspect of these
two problems which is strikingly similar; this is that there is a primary load
which is induced and remains constant or nearly constant: in the case of the
gas turbine it is the centrifugal forces developed as the engine is started
up, and which changes, usually, only moderately during flight, and in the case
of the wellhead connector it is the large clamping force exerted by the
imposition of hydraulic pressure. Superimposed on each of these is a second
load which imposes many thousands of cycles of load per major cycle: in the
case of the gas turbine the origin is vibration, and in the case of the wellhead
fitting it is the surface motion of the vessel moving the riser. The point
which is made is that, in each case, the major load moves the contact to a
particular point in $P$-$Q$-$\sigma$ space, where these symbols mean,
respectively, the normal load, shear force and differential bulk tension
arising parallel with the contact flank, and hold it there. Then, the
secondary load, which fluctuates in what we assume to be a cyclic manner,
induces changes in all three of these quantities. As the origin of the cyclic components is a solitary load, $P$, $Q$, and $\sigma$ must change in phase.

Because the assembly suffers many thousands of cycles of minor load per major cycle the problem we set
ourselves is to establish the steady-state solution to the problem as succinctly as
possible. Finding the coordinates of the contact and the coordinates of the permanent stick zone is enough to determine the maximum extent of the slip zones, and in this paper we show how to do this.

The majority of partial slip solutions employ a method in which the shear traction distribution
is viewed as the sum of that due to sliding, together with a corrective
term. The earliest and best known normal contact
solution was that found by Hertz, i.e. where the contacting bodies
have second order (strictly parabolic but usually interpreted as circular
arc) profiles \cite{Hertz_1881}, so it is natural that the first partial
slip contact solutions were all associated with the same geometry. The
first solution, for a subsequently monotonically increasing shear force, was found by
Cattaneo \cite{Cattaneo_1938}, and, apparently unaware of this solution, Mindlin
\cite{Mindlin_1949} developed the same solution and went on to look at unloading
and reloading problems \cite{Mindlin_1951}, \cite{Mindlin_1953}. These were the only significant
solutions for some time, and then Nowell and Hills \cite{Hills_1987} looked
at what happened when a bulk tension was simultaneously exerted in
one body as the shear force was gradually increased. The
next breakthrough came with the near simultaneous discovery by J\"ager \cite{Jaeger_1998} 
and Ciavarella \cite{Ciavarella_1998} that, just as the `corrective' shear traction was a
scaled form of the sliding shear traction for the Hertz case, the
same geometric similarity applies whatever the form of the contact. Major progress in solving problems involving a varying normal and shear load, and where the intention was to track out the full behaviour
as a function of time, was made in \cite{Barber_2011}. But
this calculation was restricted to $P$-$Q$ problems where the stick zone was symmetrically positioned and
only its extent as a function of time was to be found.

In this paper we shall restrict
our attention to contacts which are geometrically symmetric, but where the stick zone can be positioned anywhere within the contact edges and where the contacting bodies are elastically similar.  

In addition we shall establish an
approximate solution to the same problem using asymptotic forms. The latter is
done with the objective of demonstrating the equivalent effect of shear and bulk-tension at a contact edge. These asymptotes establish the contact edge behaviour, but, of
course, an additional feature of the problem described is that the changing
normal load causes the position of the contact edge to change, smearing out
the fretting damage, and the effects of this are yet to be established
experimentally.

\begin{figure}[t]
	\centering
	\includegraphics[scale=0.7, trim= 0 0 0 0, clip]{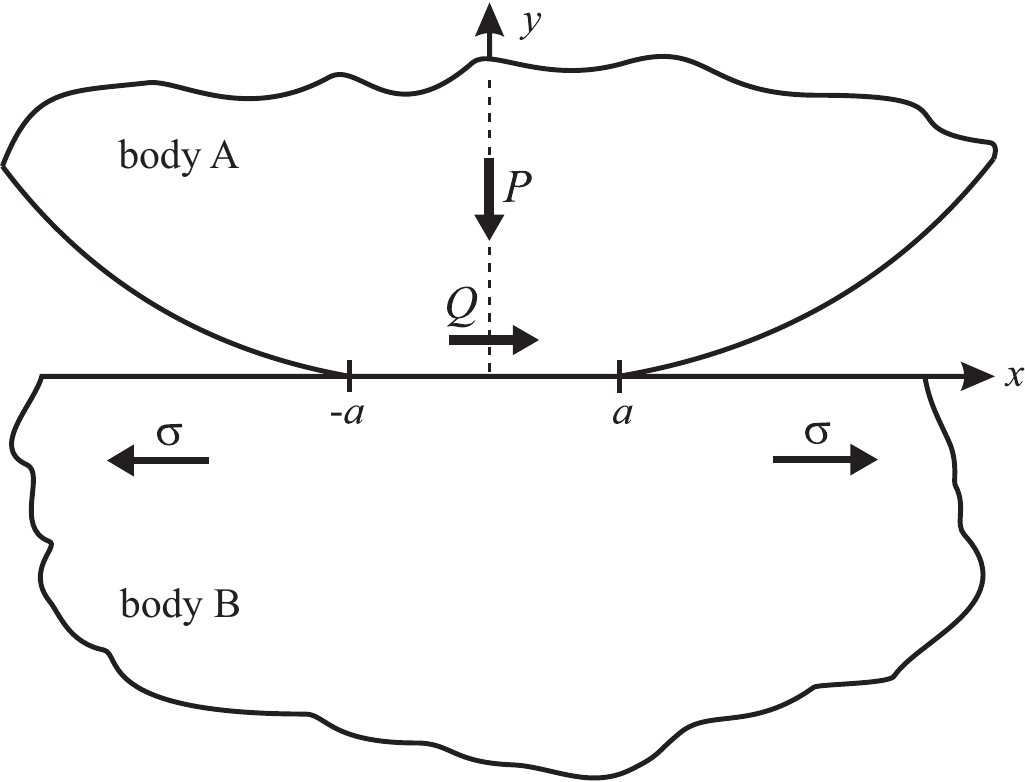}
	\caption{Generic half-plane contact subject to normal, shear and bulk tension loading}
	\label{fig:generic_half_plane}	
\end{figure} 

Lastly, we will apply the results found to a partial slip
contact whose geometry is Hertzian, simply because the algebra for this case
is most straightforward, and its extension to the more frequent practically
occurring `flat and rounded' contact case is straightforward, if algebraically taxing.
Figure \ref{fig:generic_half_plane} shows a generic half-plane contact, subject to normal, shear, and
bulk tension loading, for reference. The contact law specifies the half-width
of the contact, $a(P)$. 

\begin{figure}[H]
	\centering
	\includegraphics[scale=0.45, trim= 30 350 200 0, clip]{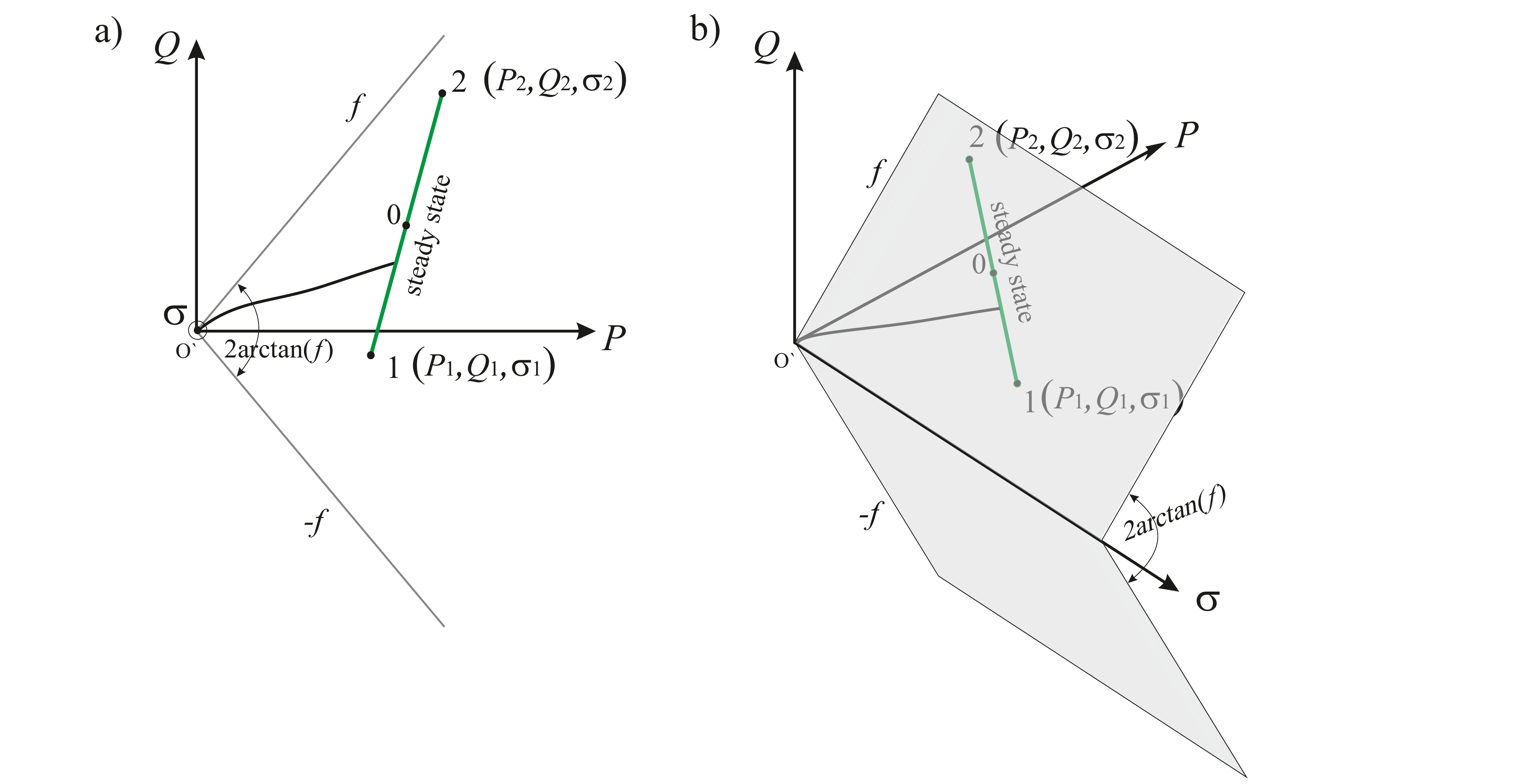}
	\caption{Two (a) and three-dimensional (b) illustration of a load space for a $P$-$Q$-$\sigma$ problem}
	\label{fig:PQsigma_load_space}	
\end{figure}

It is valuable to visualise the history of loading in the two-dimensional
load space depicted in Figure \ref{fig:PQsigma_load_space} a). The initial loading might take us to a point $0$ (i.e. $P_{0}$, $Q_{0}$, $\sigma_{0}$), or anywhere between the fluctuations of range ($\Delta P$, $\Delta Q$, $\Delta\sigma$) so that the steady state loading trajectory moves between points ($P_{1}$, $Q_{1}$, $\sigma_{1}$), where each of these
quantities is defined as the mean value \textit{less} half the range given by the delta terms, and ($P_{2}$, $Q_{2}$, $\sigma_{2}$), where each of these quantities is the mean value \textit{plus} half the
change. Note that the steady state trajectory is a straight line as $P$, $Q$, and $\sigma$ change in phase. Out of phase changes result in an elliptically shaped loading trajectory. Modest variations from in phase loading do not invalidate the presented solution provided that the slip zones are monotonically increasing at each end of the contact during each half-cycle. Figure \ref{fig:PQsigma_load_space} b) looks at a $P$-$Q$-$\sigma$ space from a rotated perspective, which includes planes with gradient $\pm f$ with respect to the normal load axis, where $f$ is the coefficient of friction. The loading
trajectory must lie within these planes - if it touches them, the contact will
slide, i.e. there will be rigid body motion. Notice that the presence of a
bulk tension affects the propensity of a contact to \emph{slip}, but not to
\emph{slide} (no rigid body motion).

We can also establish the condition for
complete adhesion - no slip at any point. This is given by the inequality
\cite{Hills_2011}
\begin{align}\label{eq1}
\frac{\left\vert \Delta Q\right\vert }{\Delta P}+\frac{\pi
	a\Delta\sigma}{4\Delta P}<f \, \text{,}
\end{align}
where we would normally think of the delta terms as differentials but, if they
are applied to the problem of proportional loading between the end points they
may be interpreted as stated earlier, and the value of $a$ is taken as its
largest value, here depicted $a_{2}$. When this inequality is not satisfied
there will be a steady state (or permanent) stick zone, and regions of
oscillatory slip.

\section{Steady-state slip behaviour}

\hspace{0.4cm} For the reduced problem when there is no variation in bulk tension, a
comprehensive analysis, including the effects of phase shift and transient
behaviour, was given in \cite{Barber_2011}. There, it was shown
that the size of the permanent stick zone is given by $a(P_{K})$ where
\begin{align}\label{PK}
P_{K}=P_{0}-\frac{\Delta Q}{2f}.
\end{align}

In this sequel we shall assume that the amount of bulk tension added-in is
small so that it is insufficient to reverse the direction of slip at either
end of the contact. Therefore, the slip direction is reversed when going
from point 1 to point 2 of the loading cycle compared with going from 2 to 1. The slip zones find their maximum extent just before the end points of the loading trajectory are reached, but slip at each end of the contact is always of the same sign.

To enable the solution to be developed in its simplest form, we write down,
first, in the spirit of the Ciavarella-J\"ager approach \cite{Ciavarella_1998}, the
solution to the \textit{normal} contact problem. Consider, first, the normal load
problem, Figure \ref{fig:generic_half_plane}. At point $i\ (i=1,2)$ the contact pressure is $p_{i}(x)
$ and contact half-width $a_{i}$. These quantities are related to the contact
profile, the gap function $g(x)$, by \cite{Barber_2010}
\begin{align}\label{gap_function}
\frac{\mathrm{d}g}{\mathrm{d}x}  &  =\frac{A}{\pi}\int_{-a_{i}}^{a_{i}}%
\frac{p_{i}(\xi)\,\mathrm{d}\xi}{x-\xi}\, \text{,}-a_{i}<x<a_{i}\,\text{,}
\end{align}
where $A=\frac{4(1-\nu^{2})}{E}$ is the material compliance of the two
half-planes, $E$ being the Young's modulus and $\nu$ Poisson's ratio, and
plane-strain obtains. Normal equilibrium is imposed by setting
\begin{align}
P_{i}  &  =\int_{-a_{i}}^{a_{i}}p_{i}(x)\mathrm{d}x\,,
\end{align}
where $P_{i}$ is the normal load.

\begin{figure}[H]
	\centering
	\includegraphics[scale=0.45, trim= 0 0 0 0, clip]{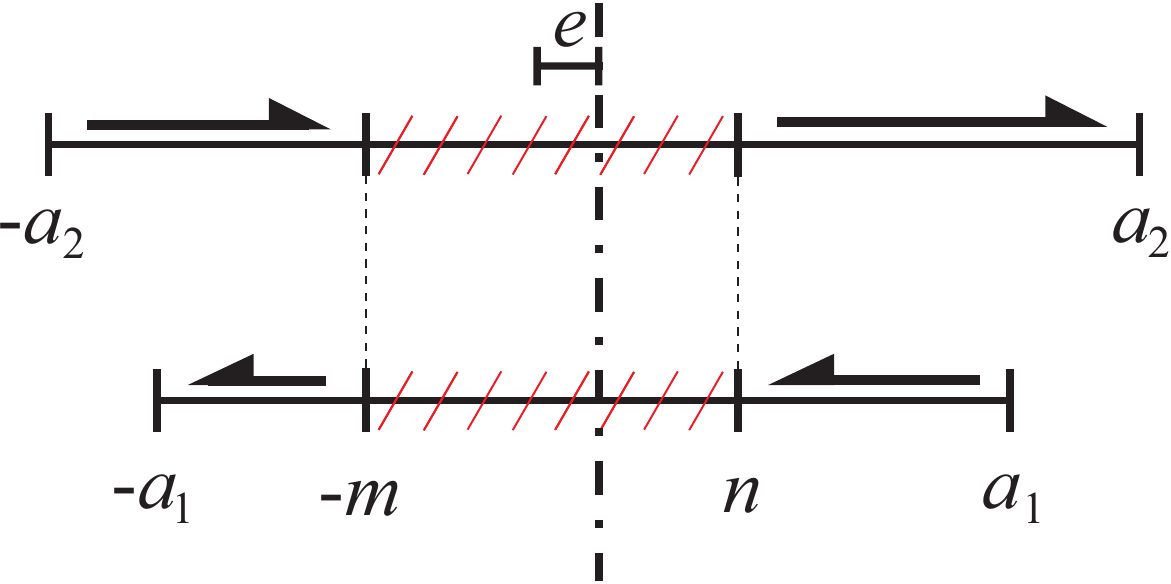}
	\caption{Contact as the ends of loading cycle are approached, including a permanent stick zone}
	\label{fig:contact_patch}	
\end{figure}

We turn, now, to tangential loading. Figure \ref{fig:contact_patch} shows, schematically, the
contact as the ends of the loading cycle are approached, including a permanent stick zone in
the steady-state stick zone, spanning $\left[-m,n\right]$, where $e$ indicates the eccentricity from the centre line of the contact. A restriction of the solution presented is that the direction of slip is the same at either end of the contact when going from 1 to 2 and is reversed, but remains the same at both ends when going from 2 to 1. This manifests in the constraint that $-a_1<-m$ and $n<a_1$, where $a_1$ is the contact half-width at the minimum point of the load cycle.  

Because of the
presence of bulk stress this stick zone is not located centrally. Hence the
slip zones at the ends of the contact are not of the same size and the problem
becomes unsymmetrical. Note that the extent of the slip zones (their inner
limits) must be the same at the end of the cycle to preserve continuity of
material: the slip displacement occurring in one half cycle must be precisely
equal and opposite to the displacement in the other half cycle for slip not to accumulate.

We know that in the stick zone, $-m<x<n$, the change in strain difference
between the two bodies, at the surface, must be zero between the two loading states
and we can write

\begin{align}\label{strain_difference}
\Delta\varepsilon_{xx,1}=\Delta\varepsilon_{xx,2}\,\text{,}
\qquad-m<x<n\text{.}
\end{align}
%\begin{align}\label{strain_difference}
%\frac{\mathrm{d}g_{1}}{\mathrm{d}x}=\frac{\mathrm{d}g_{2}}{\mathrm{d}x}\,\text{,}
%\qquad-b<x<c\text{.}
%\end{align}

The difference in surface strains arising in the two bodies, $\Delta\varepsilon_{xx,i}$, we will
find from the customary integral equation, by writing the shear traction as
the sum of the full sliding term, $fp_{i}(x)$, over the full contact length,
together with a corrective term $q_{i}^{\ast}(x)$ in the stick zone and the
effect of the difference in bulk stress. For load point \char`\"{}1\char`\"{}
we write \cite{Barber_2010}
\begin{align}
\Delta\varepsilon_{xx,1}=-\frac{A}{\pi}\int_{-a_{1}}^{a_{1}}
\frac{fp_{1}(\xi)\,\mathrm{d}\xi}{\xi-x}+\frac{A}{\pi}\int_{-m}^{n}\frac{q_{1}^{\ast}(\xi)\,\mathrm{d}\xi}{\xi-x}+\frac{A}{4}\,\sigma_{1}\text{ ,}
\end{align}
and for load point \char`\"{}2\char`\"{} we write
\begin{align}
\Delta\varepsilon_{xx,2}=\frac{A}{\pi}\int_{-a_{2}}^{a_{2}}
\frac{fp_{2}(\xi)\,\mathrm{d}\xi}{\xi-x}+\frac{A}{\pi}\int_{-m}^{n}\frac{q_{2}^{\ast}(\xi)\,\mathrm{d}\xi}{\xi-x}+\frac{A}{4}\,\sigma_{2}\text{ .}
\end{align}

We infer from equation \eqref{strain_difference} that
\begin{align}
&\qquad \qquad  -\frac{A}{\pi}\int_{-a_{1}}^{a_{1}}\frac{fp_{1}(\xi)\,\mathrm{d}\xi}{\xi-x
}+\frac{A}{\pi}\int_{-m}^{n}\frac{q_{1}^{\ast}(\xi)\,\mathrm{d}\xi}{\xi-x
}+\frac{A}{4}\,\sigma_{1}   \nonumber \\
&\qquad \qquad  =\frac{A}{\pi}\int_{-a_{2}}^{a_{2}}\frac{fp_{2}(\xi)\,\mathrm{d}\xi}{\xi-x
}+\frac{A}{\pi}\int_{-m}^{n}\frac{q_{2}^{\ast}(\xi)\,\mathrm{d}\xi}{\xi-x
}+\frac{A}{4}\,\sigma_{2}\, \text{,}\qquad-m<x<n\, \, \text{,}
\end{align}
and as $-a_{2}<-a_{1}<-m<x<n<a_{1}<a_{2}$, we may rewrite the above equation, making use of equation \eqref{gap_function},
in the usual form for singular integral equations as
\begin{align}\label{SIE_q1-q2*}
\frac{2f}{A}\frac{\mathrm{d}g}{\mathrm{d}x}+\frac{\Delta\sigma}{4}=-\frac
{1}{\pi}\int_{-m}^{n}\frac{\left[  q_{2}^{\ast}-q_{1}^{\ast}\right]
(\xi)\,\mathrm{d}\xi}{\xi-x}\text{ ,}-m<x<n\, \, \text{,}
\end{align}
where $\Delta\sigma=\sigma_{1}-\sigma_{2}$. It is worth commenting that the
LHS of this equation has just two terms; one is the profile of the contacting
bodies (as appears in the normal load problem), and the other is a constant. Let the resultant `corrective' shear forces be
\begin{align}
Q_{i}^{\ast}=\int q_{i}^{\ast}(x)\,\mathrm{d}x\, \text{,}
\end{align}
so that we may now impose tangential equilibrium by setting
\begin{align}
Q_{1}  &  =-fP_{1}+Q_{1}^{\ast}\, \text{,}\\
Q_{2}  &  =fP_{2}+Q_{2}^{\ast}\, \text{.}
\end{align}
The range of shear force, $\Delta Q$, is given by
\begin{align}
\Delta Q=Q_{2}-Q_{1}=f(P_{2}+P_{1})+\int_{-m}^{n}[q_{2}^{\ast}-q_{1}^{\ast
}](x)\,\mathrm{d}x\,\text{.}
\end{align}

\subsection{Partial solution}
\hspace{0.4cm} It is possible to make some progress towards a general solution without
specifying the contact geometry. The inversion of integral equation \eqref{SIE_q1-q2*},
bounded both ends, is given by
\begin{align}\label{SIE}
\lbrack q_{2}^{\ast}-q_{1}^{\ast}](x)=-\frac{1}{\pi}\sqrt{(x+m)(n-x)}\int%
_{-m}^{n}\frac{\left[  \frac{\Delta\sigma}{4}+\frac{2f}{A}\frac{\mathrm{d}%
		g}{\mathrm{d}\xi}\right]  \mathrm{d}\xi}{\sqrt{(\xi+m)(n-\xi)}(\xi
	-x)}\,\text{,}%
\end{align}
and the consistency condition is given by
\begin{align}
\int_{-m}^{n}\frac{\left[  \frac{\Delta\sigma}{4}+\frac{2f}{A}\frac
	{\mathrm{d}g}{\mathrm{d}\xi}\right]  \mathrm{d}\xi}{\sqrt{(\xi+m)(n-\xi)}%
}=0\,\text{.}%
\end{align}

Because of the identity $\int_{-m}^{n}\frac{1}{\sqrt{\left(  m+\xi \right)  \left(  n-\xi\right)
}}\mathrm{d}\xi=\pi$, this gives the result
\begin{align}\label{consistency_cond}
\int_{-m}^{n}\frac{\frac{\mathrm{d}g}{\mathrm{d}\xi}}{\sqrt{\left(
		m+\xi\right)  \left(  n-\xi\right)  }}\mathrm{d}\xi=-\frac{\pi A\Delta\sigma}{8f}\, \text{,}
\end{align}
which provides part of one result helping to fix the position and extent of the permanent stick region.
Further, a second mathematical identity
\begin{align}
\int_{-m}^{n}\frac{\mathrm{d}\xi}{\sqrt{(\xi+m)(n-\xi)}(\xi-x)}=0\, \text{,}\qquad-m\leq
x\leq n\, \text{,}
\end{align}
simplifies the solution of the singular integral equation \eqref{SIE} to give
\begin{align}
\lbrack q_{2}^{\ast}-q_{1}^{\ast}](x)=-\frac{2f}{\pi A}\sqrt{(x+m)(n-x)}%
\int_{-m}^{n}\frac{\frac{\mathrm{d}g}{\mathrm{d}\xi}\mathrm{d}\xi}{\sqrt
	{(\xi+m)(n-\xi)}(\xi-x)}\, \text{,}\qquad-m\leq x\leq n.
\end{align}

To evaluate the tangential equilibrium condition, the following integral is needed. If we change the order of integration we find
\begin{align}
\int_{-m}^{n}[q_{2}^{\ast}-q_{1}^{\ast}](s)\mathrm{d}s  &  =-\frac{2f}{A}\int_{-m}%
^{n}\frac{\sqrt{\left(  m+s\right)  \left(  n-s\right)  }}{\pi}\left(
\int_{-m}^{n}\frac{\frac{\mathrm{d}g}{\mathrm{d}\xi}}{\sqrt{\left(
		m+\xi\right)  \left(  n-\xi\right)  }\left(  \xi-s\right)  }\mathrm{d}%
\xi\right)  \mathrm{d}s\\
&=  -\frac{2f}{A\pi}\int_{-m}^{n}\frac{\frac{\mathrm{d}g}{\mathrm{d}\xi}}%
{\sqrt{\left(  m+\xi\right)  \left(  n-\xi\right)  }}\left(  \int_{-m}%
^{n}\frac{\sqrt{\left(  m+s\right)  \left(  n-s\right)  }}{\left(
	\xi-s\right)  }\mathrm{d}s\right)  \mathrm{d}\xi \, \text{.}
\end{align}

Knowing that
\begin{align}
\int_{-m}^{n}\frac{\sqrt{\left(  m+s\right)  \left(  n-s\right)  }}{\left(
	\xi-s\right)  }\,\mathrm{d}s=-\frac{\pi}{2}\left(  m-n\right)  +\pi\xi\, \text{,}\qquad-m\leq\xi\leq n\, \text{,}
\end{align}
we see that
\begin{align}
\int_{-m}^{n}[q_{2}^{\ast}-q_{1}^{\ast}](s)\mathrm{d}s=-\frac{2f}{A\pi}\int_{-m}%
^{n}\frac{\frac{\mathrm{d}g}{\mathrm{d}\xi}\left(  -\frac{\pi}{2}\left(
	n-m\right)  +\pi\xi\right)  }{\sqrt{\left(  m+\xi\right)  \left(
		n-\xi\right)  }}\mathrm{d}\xi\, \text{,}
\end{align}
and making use of equation \eqref{consistency_cond}, tangential equilibrium is evaluated as 
\begin{align}\label{tangential_equilibrium}
f\left(  P_{2}+P_{1}\right)  -\Delta Q-\frac{\pi\left(  n-m\right)
	\Delta\sigma}{8}=\frac{2f}{A}\int_{-m}^{n}\frac{\xi\frac{\mathrm{d}%
		g}{\mathrm{d}\xi}}{\sqrt{\left(  m+\xi\right)  \left(  n-\xi\right)  }%
}\mathrm{d}\xi\, \text{.}
\end{align}

\section{Application - Hertzian contact}

\hspace{0.4cm} In a Hertzian contact, of relative radius of curvature $R$, the contact law is
given by \cite{Hertz_1881}
\begin{align}\label{Hertz_contactlaw}
a^{2}=\frac{2PAR}{\pi} \, \text{,}
\end{align}
so that if the normal load varies over the range {[}$P_{1}\quad P_{2}]$ the half-width of the
contact patch varies over the range
\begin{align}
a_{i}^{2}=\frac{2P_{i}AR}{\pi}\qquad i=1,2 \, \text{.}
\end{align}

In addition, we know that for the present case we have $\frac{\mathrm{d}%
	g}{\mathrm{d}x}=x/R$, and so equation \eqref{tangential_equilibrium} becomes
\begin{align}
f\left(  P_{2}+P_{1}\right)  -\Delta Q-\frac{\pi\left(  n-m\right)
	\Delta\sigma}{8}=\frac{2f}{A}\frac{\pi}{8R}\left(  3m^{2}+3n^{2}-2mn\right) \, \text{.}
\end{align}

Turning to the consistency equation \eqref{consistency_cond}, the following integral appears
\begin{align}
\int_{-m}^{n}\frac{\frac{\mathrm{d}g}{\mathrm{d}\xi}}{\sqrt{\left(
		m+s\right)  \left(  n-s\right)  }}\mathrm{d}s=\frac{\pi}{2R}\left(  n-m\right)
 \, \text{.}
\end{align}

So the relation between the eccentricity, $e$, and the range of the bulk stress is given by
\begin{align}\label{Hertz_shift_stick_zone}
e=\frac{\left(  n-m\right)  }{2}=-\frac{AR\Delta\sigma}{8f} \, \text{,}
\end{align}
and using this result in the equilibrium equation leads to
\begin{align}
f\left(  P_{2}+P_{1}\right)  -\Delta Q=\frac{1}{4}\frac{\pi}{AR}f\left(
m+n\right)  ^{2} \, \text{.}
\end{align}

The size of the permanent stick region,  $d=\frac{\left(  m+n\right)  }{2}$,
is given by
\begin{align}\label{Hertz_permanent_stick_zone}
d^{2}=\frac{AR}{\pi}\left[  P_{1}+P_{2}-\frac{\Delta Q}{f}\right] \, \text{,}
\end{align}
and $n=d+e,-m=-d+e$. The size of the RH slip zone therefore varies between
$\left[  a_{2}-(d+e)\right]  $ and $\left[  a_{1}-\left(  d+e\right)  \right]
$ and the range of the LH slip zone is given by similar expressions but with
$e $ replaced by $-e$.

It is noteworthy that, in the case of a Hertzian contact, evaluation of the consistency condition and tangential equilibrium leads to a pair of uncoupled explicit equations. If the geometry profile gradient, $\frac{\mathrm{d}g}{\mathrm{d}x}$, is different from a Hertzian contact, i.e. is not a linear function, the two equations might be coupled and the size, $d$, and the eccentricity, $e$, of the permanent stick region are to be obtained implicitly.

\begin{figure}[H]
	\centering
	\includegraphics[scale=0.85, trim= 0 0 0 0, clip]{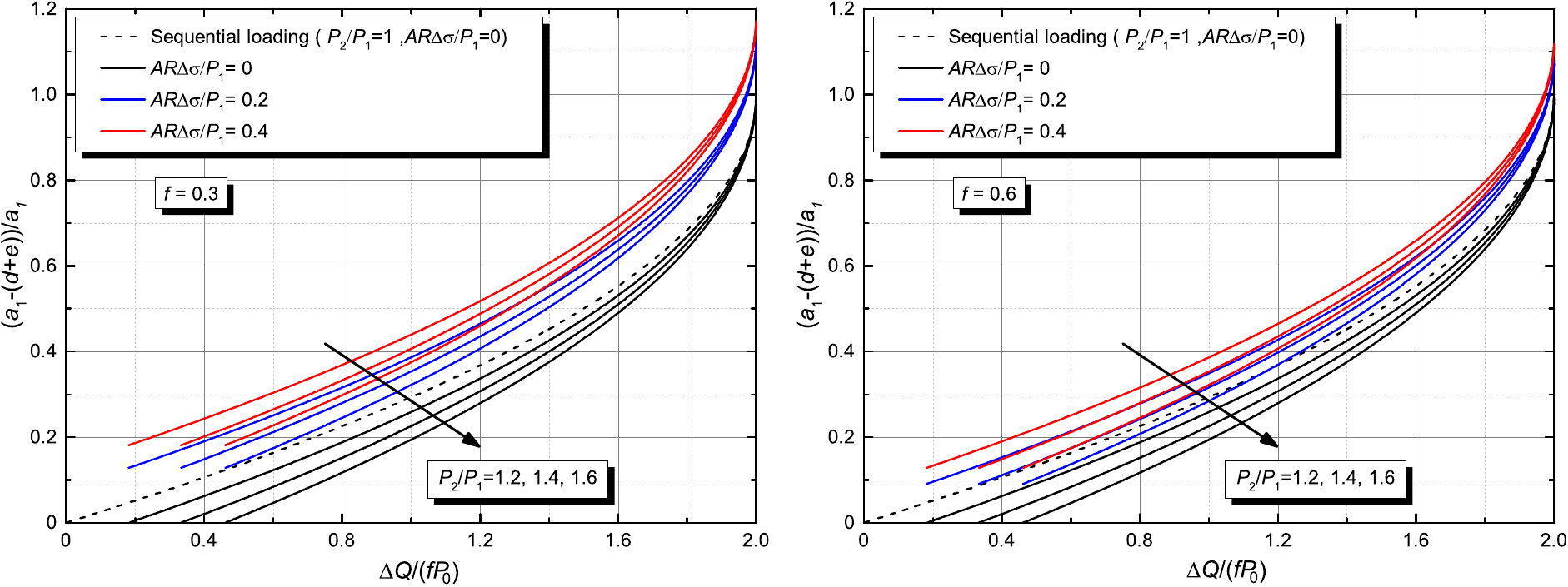}
	\caption{Evolution of slip zone sizes for different loading cases}
	\label{fig:slip_zones}	
\end{figure} 

Figure \ref{fig:slip_zones} shows the length of the
slip zone at the right edge of the contact in a dimensionless form. The length of the slip zone is represented for different load regimes: $\Delta Q/\left(
fP_{0}\right) $, $P_{2}/P_{1}$ and $\Delta \sigma /P_{1}$. We also
consider different values for the coefficient of friction $\left( f=0.3\text{%
	, }0.6\right) $. For comparison, the expected slip zone for the sequential loading case without a change in bulk stress ($%
P_{2}/P_{1}=1$, $\Delta\sigma = 0$) is included. As expected,
the greater the ratio between $\Delta Q/\left( fP_{0}\right) $, the longer is
the slip zone. A similar behaviour is found when keeping the other
loading parameters constant, but increasing the range of the bulk stress. On the other
hand, increasing the ratio $P_{2}/P_{1}$, while maintaining a constant $\Delta
Q/\left( fP_{0}\right) $ and $\Delta \sigma /P_{1}$, decreases the slip
zone. Finally, similar conclusions can be drawn for the case with $%
f=0.6$, but noting that the variation of the slip zone length with $%
\Delta \sigma /P$ is less pronounced than that obtained in the case with a
coefficient of friction equal to $0.3$.

\section{Asymptotic representation}

\begin{figure}[H]
	\centering
	\includegraphics[scale=0.7, trim= 0 0 0 0, clip]{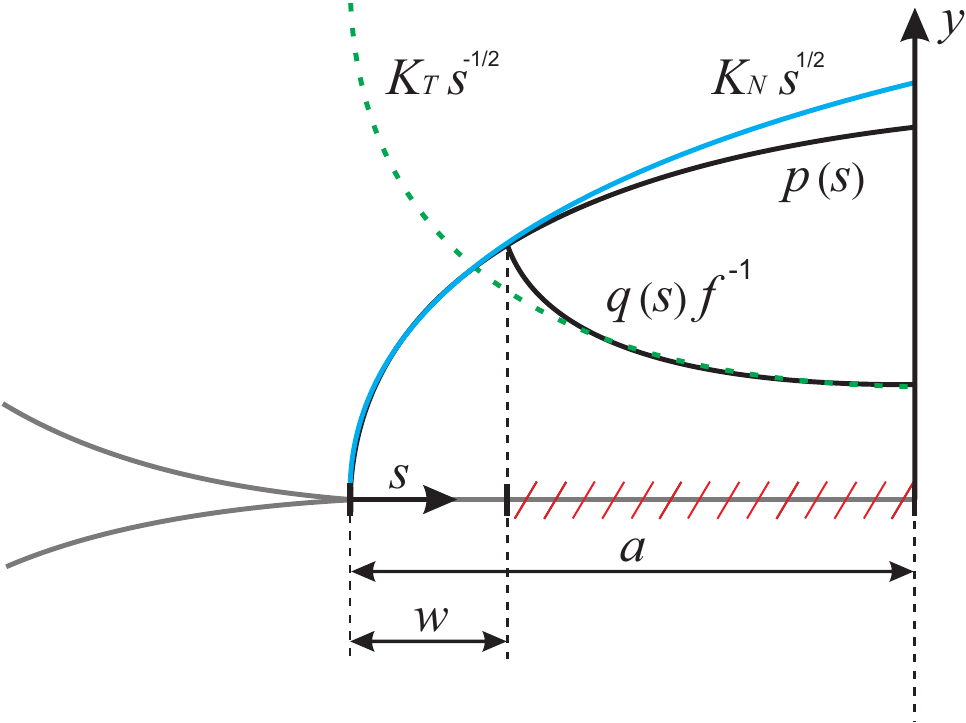}
	\caption{Normal square root bounded asymptote and tangential square root singular asymptote}
	\label{fig:edge_asymptotes}	
\end{figure}

\hspace{0.4cm} The object of this approach is to be able to provide a description of the
contact edge state of tractions and slip/stick in the smallest number of parameters, and to apply them to a range of problems. This enables a prototype
geometry to be replaced by another in the laboratory test - for example a
flat and rounded punch by a Hertzian contact - and maintaining a very similar
edge contact pressure. By changing the mix of shear to bulk tension
exciting shear tractions slip zones of similar length may be maintained. These ideas have been explored
quite extensively for problems where the normal load remains constant \cite{Hills_2016}. Note that, because we are using a single term asymptote as shown in Figure \ref{fig:edge_asymptotes}, the
solutions will be poor if the contact edge moves significantly in comparison
with the contact size. 

It is therefore likely to work well with flat and
rounded contacts, so that $a_{2}-a_{1}\ll a_{0}$ and where the change in normal load is such that $\Delta P \ll P_0$. Furthermore, the asymptotic representation hinges upon the assumption that the size of the slip zones is small in comparison with the contact size. From \cite{Fleury_2017} we see that, generally, if a normal load has
already been applied and held constant, the normal traction might be approximated as $p(s)=K_N \sqrt{s}$, where the stress intensity factor $K_N$ is connected to the contact law by 
\begin{align}
K_{N}=\frac{1}{\pi} \sqrt{\frac{2}{a}} \frac{\mathrm{d}P}{\mathrm{d}a}\, \text{.}
\end{align}

If for the generated contact of half width, $a$, the coefficient of friction is sufficiently high to inhibit all slip,
the application of a shear force, $Q$, and bulk tension, $\sigma$, will induce, at
the left hand edge ($x=-a$) a singular shear traction whose form is given by
$q$$\left(  s\right)= K_{T}/\sqrt{s}$, where $x=-a+s\,$$ \text{ for } \, \, s\ll a$, and where
\begin{align}
K_{T}=\frac{Q}{\pi\sqrt{2a}}+\frac{\sigma}{4}\sqrt{\frac{a}{2}}\, \text{,}
\end{align}
and positive values of $Q$ and $\sigma$ are as depicted in Figure \ref{fig:PQsigma_load_space}. Let an equivalent
shear force alone, say $S$, produce the same value of stress intensity,
$K_{T}.$ Then, clearly
\begin{align}\label{equivalent_shear_force}
S=Q\pm\frac{\pi \,a\,\sigma}{4} \, \text{,}
\end{align}
where we choose the $+ve$ sign for the LH end where the effects of shear and
bulk tension add, and the $-ve$ sign for the RH end where the bulk tension
diminishes the effect of the shear force.

So, the size of the permanent stick zone estimated by this procedure is given by
equation \eqref{PK} with $Q$ replaced by $S$ so that our best estimate of the
stick-slip transition points, $-m=a(P_{K})$, $n=a(P_{K})$ is given by
\begin{align}
P_{K}=P_{0}-\frac{\Delta S}{2f} \, \text{,}
\end{align}
where, for the former we choose the $+ve$ sign in equation \eqref{equivalent_shear_force} and for the latter the $-ve$ sign.

\begin{figure}[H]
	\centering
	\includegraphics[scale=0.5, trim= 0 0 0 0, clip]{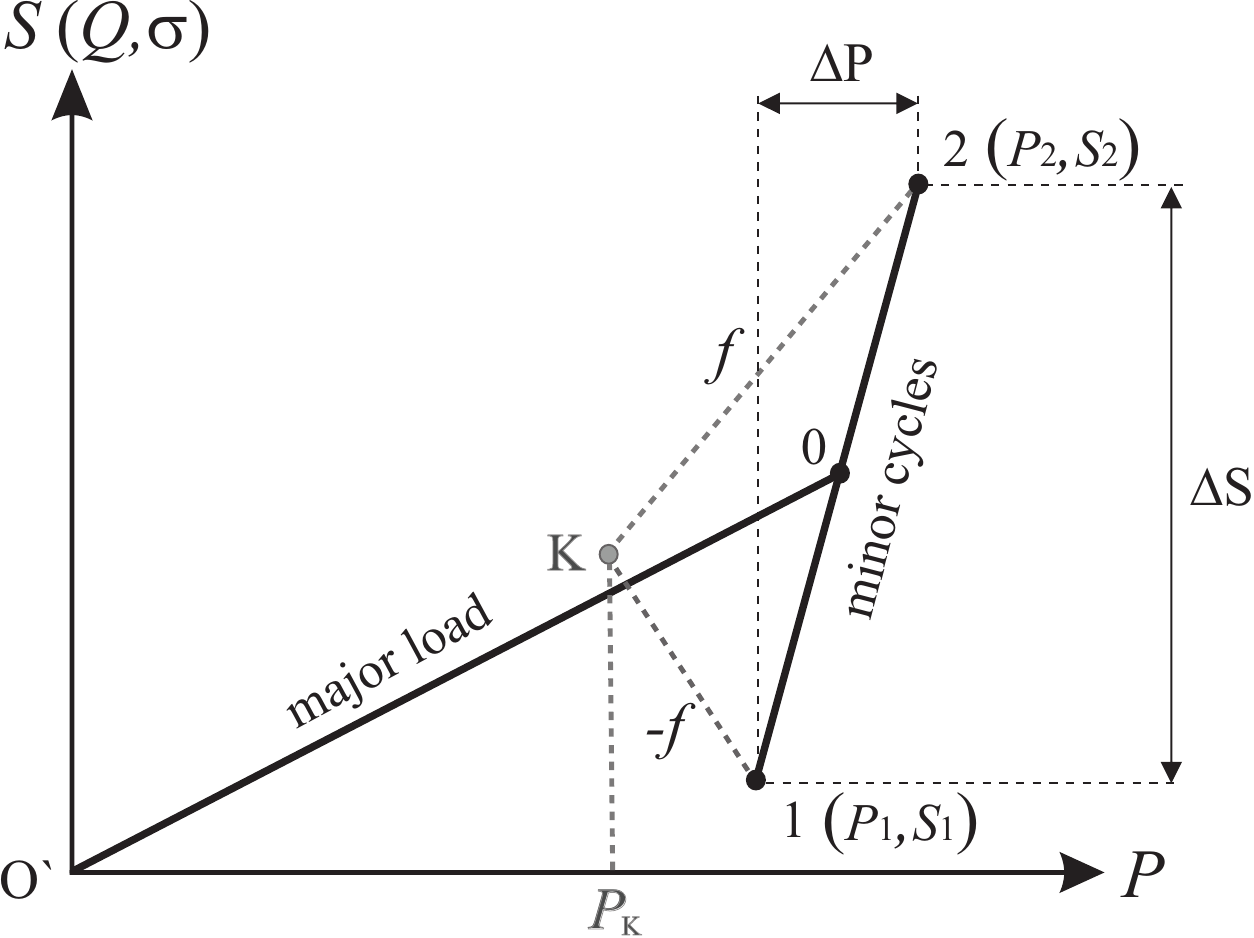}
	\caption{Two-dimensional $P$-$S$ load space for a $P$-$Q$-$\sigma$ problem}
	\label{fig:PS_load_space}	
\end{figure}

The equivalent shear force, $S$, collapses the three-dimensional $P$-$Q$-$\sigma$ problem to two dimensions as depicted in Figure \ref{fig:PS_load_space}. 
Obviously this facilitates visualisation. However, note that $P$-$S$ load space represents partial slip behaviour at one of the edges of the contact only. So two loops (or straight lines depending on the phase shift between the load components) are needed to describe fully the contact's partial slip behaviour at both edges in the collapsed two-dimensional load space.

An asymptotic formulation for incomplete contacts subject to varying normal and shear loads has been developed in \cite{Fleury_2017}. It is possible to extend these methods to problems involving differential bulk stresses. Here, we wish to demonstrate that the mathematically exact results we obtain for the steady-state slip behaviour in this paper agree very well with the asymptotic solution found for a Hertzian contact. 

For brevity, the findings from \cite{Fleury_2017} are not repeated in detail, but the steady-state behaviour may be described as follows. The size of the slip zone, $w= a-d \pm e$, as we reach point 2 when loading from 1 to 2 in steady-state, may be approximated by
\begin{align}
w_2 &= -\frac{\Delta K_T}{f K_N^2}+\frac{\Delta a_{12}}{2}\, \text{ ,}
\end{align}
where $\Delta K_T =K_T^1-K_T^2$ is the change in shear stress intensity factor $K_T$ and $\Delta a_{12} = a_1 - a_2$ is the change in contact size between minimum and maximum steady-state loading points, 1 and 2. $K_N^2$ is the normal stress intensity factor at load point 2.

The approximate size of the slip zone at the minimum loading point 1 during steady-state regime is given by
\begin{align}
w_1 &= \frac{\Delta K_T}{f K_N^1}-\frac{\Delta a_{12}}{2}\, \text{ ,}
\end{align}
where $K_N^1$ is the normal stress intensity factor at load point 1.

The approximate size of the slip zone at a general point $X$, while loading from point 1 to point 2 during steady-state regime, is given by
\begin{align}\label{forward_slip}
w_{1X} &= \frac{\Delta K_T^{1X}}{f K_N^X}-\frac{\Delta a_{1X}}{2} \, \text{ ,}
\end{align}
where $\Delta K_T^{1X} =K_T^1-K_T^X$ is the change in shear stress intensity factor $K_T$ between point 1 and $X$ and $\Delta a_{1X} = a_1 - a_X$ is the change in contact size between the minimum of the steady-state cycle and a general loading point $X$ along the steady-state trajectory.  $K_N^X$ is the normal stress intensity factor at the general load point $X$.
The approximate size of the slip zone at a general point $X$, while unloading from point 2 to point 1 during steady-state regime, is given by
\begin{align}\label{reverse_slip}
w_{2X} &= \frac{\Delta K_T^{2X}}{f K_N^X}-\frac{\Delta a_{2X}}{2}\, \text{,}
\end{align}
where $\Delta K_T^{2X} =K_T^2-K_T^X$ is the change in stress intensity factors $K_T$ and $\Delta a_{1X} = a_1 - a_X$ is the change in contact size between the maximum of the steady-state cycle and a general loading point $X$ along the steady-state trajectory.

\begin{figure}[H]
	\centering
	\includegraphics[scale=0.7, trim= 0 0 0 0, clip]{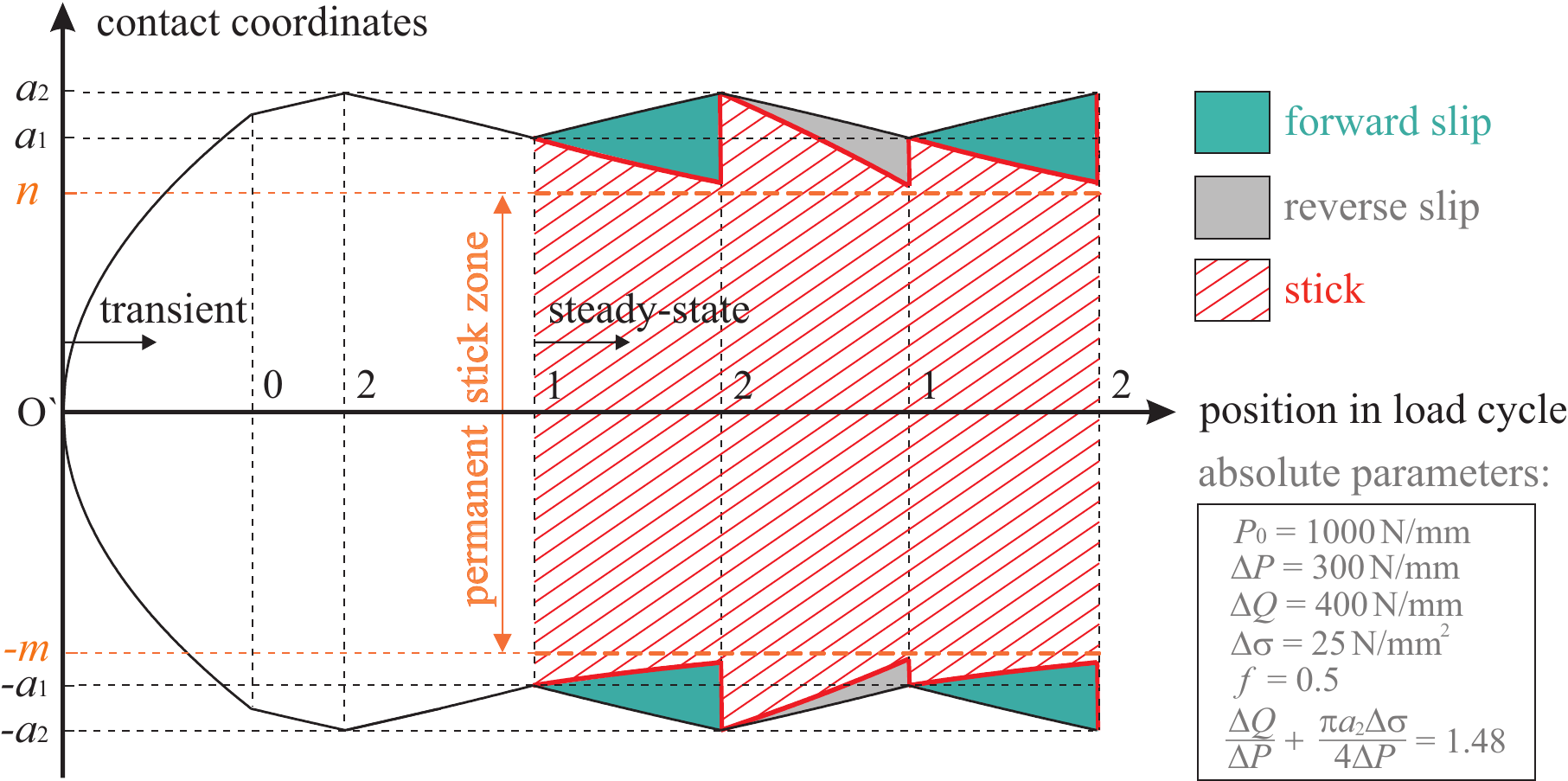}
	\caption{Contact and stick zone sizes based on analytical and asymptotic prediction for a Hertzian contact}
	\label{fig:slip_behaviour}	
\end{figure}

Figure \ref{fig:slip_behaviour} shows the slip-stick behaviour during steady-state for an example Hertzian contact during steady-state. The permanent stick zone is given by the analytical description introduced in this paper and the evolving size of the slip zones is predicted asymptotically. As there does not exist any analytical solution for the evolving slip zones in the presence of bulk tension, the asymptotic error can only be estimated based on the maximum extent of the slip zones as the extremes of the loading cycle are approached. From \cite{Fleury_2017} we know that for a varying $P$-$Q$ problem the error for the asymptotic estimation of the slip zone extent remains smaller than $10\%$ for our loading parameters. We refrain from further assessing the absolute or relative error between asymptotic or analytical solution for the plotted example as accuracy of the solution will vary significantly for different loading scenarios. The parameters were deliberately chosen so as to make the qualitative evolution of the slip zone size most visual, which strains the quantitative accuracy of the asymptotic solution. 

The contact edge behaviour is calculated by the contact law, given in equation \eqref{Hertz_contactlaw}, and the normal force, $P$, is assumed to vary linearly in a triangular waveform. The permanent stick zone is given by equations \eqref{Hertz_permanent_stick_zone} and \eqref{Hertz_shift_stick_zone}. The evolving stick-slip boundaries are estimated from equations \eqref{forward_slip} and \eqref{reverse_slip}. The solution enters the steady state when it reaches load point $1$ for the first time. At point $1^+$ the contact is fully adhered, where the \textit{plus} indicates the load state when one end of the loading path has just been passed and a \textit{minus} indicates the point when a change in sign of the loading increment is imminent. A full loading cycle can then be described in four distinctive steps $1^+ \rightarrow 2^- \rightarrow 2^+ \rightarrow 1^- \rightarrow 1^+$. As the load path approaches load point $2^-$, the material in contact outside the permanent stick zone experiences forward slip at both ends of the contact. At point $2^+$ the contact becomes fully adhered again. As the load trajectory leads back to point $1^-$ surface particles experience slip in the reverse direction so that when load point $1^-$ is reached the material within the slip region arrives at its original location where it left from at load point $1^+$. This process repeats itself every loading cycle so that potentially severe fretting damage is accumulated in the slip regions of an incomplete contact.

\section{Conclusions}
 \hspace{0.4cm} A simple set of results for obtaining the steady-state stick zone - \textit{size} and \textit{position} - for half-plane contacts, avoiding the complication of the marching in time procedure, has been set out. The results are independent of the indenter profile outside of the permanent stick zone. When the permanent stick zone and the contact law are known, the maximum extent of the slip region at either end of the contact is also effectively established. The findings are applied to a Hertzian contact and simple algebraic expressions are presented. We note that there is a striking similarity between proportional and sequential loading when applying the analytical procedure presented for obtaining a steady-state stick zone. Lastly, an  asymptotic procedure is applied to the problem. It provides an approximate solution for the marching in time procedure, and is particularly attractive when the contact law is unknown or needlessly difficult to obtain. The solutions are applied to a Hertzian contact and, at the end points of the load cycle, a comparison with the analytical results is given. 
 %This similarity might be exploited in order to facilitate fretting fatigue experiments, which try to mimic changes in contact size in addition to changes in tangential loading. 

\section*{Acknowledgments}
 \hspace{0.4cm} This project has received funding from the European Union's Horizon 2020 research and innovation programme under the Marie Sklodowska-Curie agreement No 721865. David Hills thanks Rolls-Royce plc and the EPSRC for the support under the Prosperity Partnership Grant “Cornerstone: Mechanical Engineering Science to Enable Aero Propulsion Futures”, Grant Ref: EP/R004951/1.

\bibliographystyle{unsrt}

\newpage

\end{document}